\documentclass[fleqn,11pt,twoside]{article}

\usepackage{amsthm,amsthm,amssymb, color, xcolor,epsfig, graphics, 
}

\usepackage{amsmath, graphicx, latexsym, lscape}

\makeatletter
\newcommand{\copyrightnote}[2]{{\renewcommand{\thefootnote}{}
 \footnotetext{\small\it
\begin{flushleft}
 \copyright \ #1   #2  
\end{flushleft}}}}

\newcommand{\Name}[1]{\begin{flushleft}
                       \LARGE \bf #1
                       \end{flushleft}\vspace{-3mm}}

\newcommand{\Author}[1]{\begin{flushleft}
                       \it #1 \end{flushleft}}

\newcommand{\Address}[1]{\begin{flushleft}
                       \it #1 \end{flushleft}}

\newcommand{\Date}[1]{\begin{flushleft}
                      \small  \it #1 \end{flushleft}}

\newcommand{\evenhead}{Author \ name}
\newcommand{\oddhead}{Article \ name}

\renewcommand{\@evenhead}{
\hspace*{-3pt}\raisebox{-15pt}[\headheight][0pt]{\vbox{\hbox to \textwidth
{\thepage \hfil \evenhead}\vskip4pt \hrule}}}
\renewcommand{\@oddhead}{
\hspace*{-3pt}\raisebox{-15pt}[\headheight][0pt]{\vbox{\hbox to \textwidth
{\oddhead \hfil \thepage}\vskip4pt\hrule}}}
\renewcommand{\@evenfoot}{}
\renewcommand{\@oddfoot}{}

\long\def\@makecaption#1#2{%
  \vskip\abovecaptionskip
  \sbox\@tempboxa{\small \textbf{#1.}\ \ #2}%
  \ifdim \wd\@tempboxa >\hsize
    {\small \textbf{#1.}\ \ #2}\par
  \else
    \global \@minipagefalse
    \hb@xt@\hsize{\hfil\box\@tempboxa\hfil}%
  \fi
  \vskip\belowcaptionskip}

\newcommand{\JNMPnumberwithin}[3][\arabic]{%
  \@ifundefined{c@#2}{\@nocounterr{#2}}{%
    \@ifundefined{c@#3}{\@nocnterr{#3}}{%
      \@addtoreset{#2}{#3}%
      \@xp\xdef\csname the#2\endcsname{%
        \@xp\@nx\csname the#3\endcsname .\@nx#1{#2}}}}%
}

\newcommand{\resetfootnoterule} {
  \renewcommand\footnoterule{%
  \kern-3\p@
  \hrule\@width.4\columnwidth
  \kern2.6\p@}
}

\renewcommand{\footnoterule}{}

\makeatother

\theoremstyle{definition}
\newtheorem{definition}{Definition}
\newtheorem{corollary}{Corollary}
\newtheorem{remark}{Remark}

\begin{document}

\renewcommand{\evenhead}{ {\LARGE\textcolor{blue!10!black!40!green}{{\sf \ \ \ ]ocnmp[}}}\strut\hfill M A Rodr\'{\i}guez and P Tempesta}
\renewcommand{\oddhead}{ {\LARGE\textcolor{blue!10!black!40!green}{{\sf ]ocnmp[}}}\ \ \ \ \  A new discretization of the Euler equation}

\thispagestyle{empty}
\newcommand{\FistPageHead}[3]{
\begin{flushleft}
\raisebox{8mm}[0pt][0pt]
{\footnotesize \sf
\parbox{150mm}{{Open Communications in Nonlinear Mathematical Physics}\ \ \ \ {\LARGE\textcolor{blue!10!black!40!green}{]ocnmp[}}
\quad Special Issue 1, 2024\ \  pp
#2\hfill {\sc #3}}}\vspace{-13mm}
\end{flushleft}}

\FistPageHead{1}{\pageref{firstpage}--\pageref{lastpage}}{ \ \ }

\strut\hfill

\strut\hfill

\copyrightnote{The author(s). Distributed under a Creative Commons Attribution 4.0 International License}

\begin{center}
{  {\bf This article is part of an OCNMP Special Issue\\ 
\smallskip
in Memory of Professor Decio Levi}}
\end{center}

\smallskip

\Name{A new discretization of the Euler equation via the finite operator theory}

\begin{center}
\textit{In memoriam of our friend and colleague Decio Levi}
\end{center}

\Author{Miguel A. Rodr\'{\i}guez$^{\,1}$ and Piergiulio Tempesta$^{\,2}$}

\Address{$^{1}$ Departamento de F\'{\i}sica Te\'{o}rica, Facultad de Ciencias F\'{\i}sicas, Pza.~de las Ciencias 1,
  Universidad Complutense de Madrid, 28040 -- Madrid, Spain; {\tt rodrigue@ucm.es} \\[2mm]
$^{2}$ Departamento de F\'{\i}sica Te\'{o}rica, Facultad de Ciencias F\'{\i}sicas, Pza.~de las Ciencias 1,
  Universidad Complutense de Madrid, 28040 -- Madrid, Spain \\ and Instituto de Ciencias
  Matem\'aticas, c/ Nicol\'as Cabrera, No 13--15, 28049 -- Madrid, Spain; {\tt p.tempesta@fis.ucm.es}}

\Date{Received September 9, 2023; Accepted January 11, 2024}

\setcounter{equation}{0}

\begin{abstract}

\noindent 
We propose a novel discretization procedure for the classical Euler equation based on the theory of Galois differential algebras and the finite operator calculus developed by G.C. Rota and collaborators. This procedure allows us to define algorithmically a new discrete model that inherits from the continuous Euler equation a class of exact solutions.
\end{abstract}

\label{firstpage}


\section{Introduction}\label{sec:intro}

The study of discrete integrable models  over the past decades has become one of the most remarkable research lines in Mathematical Physics.

More specifically, the problem of discretizing dynamical systems in such a way that their symmetry and integrability properties are preserved has been widely investigated (see, e.g., the monographs  \cite{BS,LNP,LWY,Suris}  and the references therein). This field is one of those in which the scientific contribution of Decio Levi has been fundamental \cite{LWY}. 
His beautiful insight into the study of the multiple facets of discrete mathematics and his tireless scientific activity during several decades have influenced several generations of researchers in mathematical physics - and, in particular, the authors of this article - in a deep way.

Among the many approaches proposed for discretizing integrable models, in this paper, we shall focus  on a specific research line where  the contribution of Decio Levi has been very relevant. The underlying methodology relies on the \textit{finite operator theory} (also called Umbral Calculus), developed by G. C. Rota and his collaborators \cite{Roman,RR,Rota}. In essence, it consists of requiring the Heisenberg-Weyl algebra to be preserved in the discretization process; this, in turn, allows one to preserve the Lorentz and Galilei invariance on lattices. This technique has been introduced  in \cite{DMS} for discretizing quantum mechanics and further developed in \cite{LNO,LTW1,LT2011,LTW2}.  An application to the multiple-scale analysis of dynamical systems on a lattice has been proposed in \cite{LT2011}.

A related approach is based on the idea of preserving the Leibniz rule on lattices. This idea, under different perspectives, has been proposed in \cite{BF,Ward,Ismail,PTJDE,PTprep}. A crucial aspect of this methodology is that the standard pointwise product of functions on a lattice is replaced by a suitable nonlocal product, in such a way that discrete derivatives act on functions (or, more generally, on formal power series) as standard derivations. This framework allows one to define in  a natural way discrete versions of continuous ODEs (also in the nonlinear case), which share with them the class of exact solutions expressed in terms of power series.

Another important aspect is that by choosing
 the lattice points as the zeroes of the basic polynomials associated with the discrete derivative we are using, one can avoid the main drawback of the umbral approach to difference equations, namely the appearance of divergent power series for the solutions of the discrete models. Indeed, the series expressing the exact solutions are, by construction, truncated and thus convergent in an obvious sense. In \cite{PTprep}, this procedure has been generalized to the case of variable coefficients ODEs.

In the following, we shall adopt this theoretical framework to introduce a difference equation that we can interpret as a novel,  discrete version of the Euler equation. Indeed, under suitable hypotheses, it can be exactly solved; besides, its solutions come in a direct way from those of the standard Euler differential equation.

In this article, for the sake of simplicity, we shall keep the mathematical formalism to a minimum. A more formal approach and further technical details can be found in \cite{PTJDE,PTprep}; however, we will discuss our results in a self-contained way.

The paper is organized as follows. In Section \ref{sec2},  we review some basic notions of finite operator theory. In Section \ref{sec3}, we implement our methodology for discretizing the Euler equation. In Section \ref{sec4}, we derive some exact solutions of our discrete Euler equation and address some open questions in the final Section \ref{sec5}.

\section{The finite operator calculus: Background and notation}
\label{sec2}
The finite operator theory was proposed in \cite{Rota} and further developed in \cite{RR}, \cite{Roman} as a formal calculus adapted to treat combinatorial problems and to discuss the properties of relevant classes of polynomial sequences. It is the modern version of the umbral calculus introduced by Sylvester, Cayley, Blissard, and other authors \cite{BL2000}. In this section, we shall review some basic definitions and fundamental results of Rota's approach to difference operators.

\subsection{Delta operators and basic polynomials}
Let $\mathbb{K}$ be a field of characteristic zero and   $\mathcal{P}$ be the space of polynomials in one variable $x\in \mathbb{K}$. Let $\mathbb{N}$ be the set of non-negative integers. The operator $T: \mathcal{P} \to \mathcal{P}$ defined by $T p(x)= p(x+h)$, where $h>0$, will be said to be the \textit{shift operator}. For simplicity, we shall restrict to the case $\mathbb{K}=\mathbb{R}$. The notion of delta operators is a fundamental piece of our discretization procedure.

\begin{definition}\label{deltaop}
An operator $S$  is said to be {\it shift-invariant} if it
commutes with the shift operator $T$. A shift-invariant operator $Q$ is  a \textit{delta operator} if $Q \, x=const\neq0$.
\end{definition}
As a simple consequence of the previous definition, we have the following
\begin{corollary}\label{cor1}
For every constant $c\in \mathbb{R}$, $Q \, c=0$.
\end{corollary}
Delta operators very commonly used in the applications are  the standard derivative $D$, the forward discrete derivative $\Delta^{+}=T-\mathbf{1}$, the backward derivative $\Delta^{-}=\mathbf{1}-T$ and the symmetric derivative $\Delta^{s}=\frac{T-T^{-1}}{2}$. Many other relevant examples can be found in the literature (e.g., in \cite{Rota}).

\begin{definition}\label{def2}
 A polynomial sequence $\{p_{n}\left(x\right)\}_{n\in\mathbb{N}}$ is said to be the sequence of \textit{basic polynomials} associated with a delta operator $Q$ if the following conditions are satisfied:
\begin{align}
&  1)\text{ \ }p_{0}\left(  x\right)  =1;\notag\\
&  2)\text{ \ }p_{n}\left(
0\right)  =0\ \text{for all }n>0;\text{ \ }\nonumber\\
&  3)\text{ \ }Q p_{n}\left(  x\right)  =np_{n-1}\left(  x\right).\nonumber
\end{align}
\end{definition}
One can prove that given a delta operator $Q$, there exists a unique sequence of  basic polynomials associated with it.
In the case of the continuous derivative operator, the sequence of basic polynomials is given by $p_n(x)=x^n$. For the forward and backward discrete derivatives $\Delta^{\pm}$, the corresponding basic polynomials  are, respectively
\begin{equation}
p_{0}^{\mp}(x)=1, \hspace{2mm} p_n^{\mp}(x):=x(x\pm 1)(x \pm 2 )...(x\pm (n-1)) \label{3.7}.
\end{equation}
The basic polynomials for $\Delta^{s}$ have been explicitly determined in \cite{LT2011}. A general procedure for constructing the basic polynomials associated with an arbitrary delta operator  has been discussed in \cite{LTW2}. For completeness, we shall sketch here the salient aspects of this construction.
Let $\mathcal{A}$ be the algebra of shift-invariant operators endowed with the usual operations: the sum of
two operators, the product of a scalar with an operator, and the product of two operators. Let $x: p(x) \to x p(x)$ denote the multiplication
operator. The \textit{Pincherle derivative} of an operator $U\in \mathcal{A}$ is defined by the relation
\begin{equation}\label{PD}
U':=[U,x].
\end{equation}
In \cite{Rota}, it has been proved that if $Q$ is a delta operator, then $Q'$ is invertible. Let us introduce a conjugate operator $\beta\in \mathcal{A}$ such that the Heisenberg-Weyl algebra is satisfied \cite{LNO}:
\begin{equation}
[Q, x\beta]= \mathbf{1}.
\end{equation}
Thus, the operator $\beta$ is uniquely determined by the relation $\beta= (U')^{-1}$. To illustrate three simple examples, we recall that in the continuous case, $Q=D$, $\beta= \mathbf{1}$; for $Q=\Delta^{+}$, $\beta= T^{-1}$; for $Q= \Delta^{-}$, $\beta=T$. Generally speaking, we can formally compute the basic polynomials associated with a delta operator $Q$  through the relation 
\begin{equation}
p_n(x)= (x\beta)^n\cdot 1 \ .
\end{equation}
In particular, this relation reproduces eq. \eqref{3.7} for the case of the forward and backward derivatives.

Let $\mathcal{F}$ be the algebra of formal power series in $x$. Since the polynomials $\{p_n(x)\}_{n\in\mathbb{N}}$ for every choice of $Q$ provide a basis of $\mathcal{F}$, any $f\in\mathcal{F}$ can be expanded into a formal series of the form $f(x)=\sum_{n=0}^{\infty}a_n p_n(x) \label{exp}.$

\subsection{A general discretization scheme}
Our discretization method starts from the \textit{umbral correspondence} discussed in \cite{Rota,DMS,LTW1,LTW2}: given an ODE,  by replacing the continuous derivative with a delta operator $Q$, and the basic sequence $x^{n}$ with the basic polynomials $p_n(x)$ for $Q$, one can construct a formal discretization of the original ODE.  
However, for this discretization to be effective, it is also crucial to replace the pointwise product of functions with a suitable $*$ product, which is associative and commutative, in such a way that 
\begin{equation}
p_n(x)*p_m(x):=p_{n+m}(x). \label{starproduct}
\end{equation}
This product for the forward difference operator $\Delta$ has been proposed in \cite{Ward} and in \cite{Ismail}. Taking into account these ideas, for any $Q$, one can define the notion of \textit{Rota algebra}, introduced in \cite{PTJDE}: it is the space  $(\mathcal{F}, +,\cdot, *_{Q})$, endowed with the composition laws of sum of series, multiplication by a scalar and the $*$ product defined by eq. \eqref{starproduct}. 

If $f$ and $g$ are formal series defined on a lattice of points $\mathcal{L}$ and expanded in the polynomial basis given by a basic sequence, we have that
\begin{equation}\label{der}
\Delta (f*g) = (\Delta f)*g + f*(\Delta g)  \ .
\end{equation}
In other words, $\Delta$ acts as a derivation with respect to the $*$-product: the Leibniz rule is restored on $\mathcal{F}$. Besides, if the zeroes of the basic polynomials coincide with the set of points of our lattice $\mathcal{L}$, then we are led to an effective discretization since all the series involved truncate. We shall explain this crucial point in detail in the forthcoming discussion.

\begin{remark}
The difference equations obtained by following this procedure inherit naturally from their continuous counterparts the class of real analytic exact solutions, which are expressed in terms of convergent power series. Indeed, their discrete versions are represented in terms of finite series obtained replacing $x^n$ with the corresponding basic polynomial $p_n(x)$; the coefficients of these expansions coincide with those of the analytic solutions of the original continuous models \cite{Ward,PTJDE,PTprep}. In this technical sense, we say that this procedure preserves integrability.
\end{remark}

\section{The forward discrete derivative and the associated calculus} \label{sec3}

In the forthcoming discussion, we shall focus on the operator $\Delta^{+}:=\Delta$. To fix the notation, we propose here some useful formulas. First, we introduce a uniform lattice $\mathcal{L}$ in the positive semi-axis $\mathbb{R}_{+}$,
\begin{equation}
x_n=nh,\quad h>0, \quad n\in\mathbb{N},
\end{equation}
and, given a function $u(x)$ of a real variable, we define its discretization as the set of all of its values at the lattice points:
\begin{equation}
u_n:=u(x_n)=u(nh).
\end{equation}
The $k$-degree basic polynomials for the forward discrete derivative operator on the uniform lattice will be denoted by
\begin{equation}
(x)_0:=1,\quad (x)_k:=\prod_{j=0}^{k-1}(x-jh),\quad k=1,2,\ldots
\end{equation}
These polynomials have zeros at the lattice points $x_i$, $i=0,\ldots,k-1$. For $n\ge k$, we have:
\begin{equation}\label{pol}
(nh)_k=\prod_{j=0}^{k-1}(nh-jh)=h^kn(n-1)\cdots(n-k+1)=\frac{n!h^k}{(n-k)!},\quad n\ge k.
\end{equation}
Let $Q$ be a delta operator acting on $\mathcal{P}$, and $\{p_n(x)\}_{n\in\mathbb{N}}$ be the  basic sequence of polynomials of order $n$ associated with $Q$.  Let $\mathcal{F}$ be the algebra of formal power series in $x$. Since the polynomials $\{p_n(x)\}_{n\in\mathbb{N}}$ provide a basis of $\mathcal{F}$, they allow us to expand a  function $u(x)$ in terms of a formal power series \cite{PTJDE}:
\begin{equation}\label{FPE}
u(x)=\sum_{k=0}^{\infty}\zeta_kp_k(x),\quad \zeta_k\in\mathbb{R}.
\end{equation}
In this way, we extend the action of delta operators on functions. In general, the formal series \eqref{FPE} is divergent. However, in the case $Q=\Delta$, since  the lattice points coincide with the zeroes of $(x)_n$, the series expression for $u(x)$ on the lattice truncates. Indeed, we have:
\begin{equation}\label{un}
u_n=\sum_{k=0}^{\infty}\zeta_kp_k(nh)=\sum_{k=0}^{n}\frac{n!h^k}{(n-k)!}\zeta_k \ .
\end{equation}

We can deduce the coefficients $\zeta_k$ from the values of $u_n$ by solving a system of linear equations \cite{PTJDE}. We obtain the formula 
\begin{equation}\label{zeta}
\zeta_k=\frac{1}{h^k}\sum_{j=0}^{k}\frac{(-1)^{k-j}}{j!(k-j)!}u_j \ .
\end{equation}

In the discretization process, we replace the continuous derivatives by the operators:
\begin{equation}
u'\to \Delta u_n=\frac{1}{h}\left(u_{n+1}-u_{n}\right)
,\quad u''\to \Delta^2 u_n=\frac{1}{h^2}\left(u_{n+2}-2u_{n+1}+u_n\right),
\end{equation}
etc. The discretization of the derivative at the lattice points (the upper limit in the sums can be taken equal to $n$, because $p_i(mh)=0$ when $i>m$) can be written in terms of the basic polynomials as follows\footnote{The bounds in the sums can be adjusted so that vanishing terms are not present in the sums; we always take into account that a negative factorial in the denominator yields 0 in the corresponding term.} :
\begin{align}
u'(nh)\to& \sum_{j=0}^{n+1}j\zeta_jp_{j-1}(nh),\label{firstder}\\
u''(nh)\to& \sum_{j=0}^{n+2}j(j-1)\zeta_jp_{j-2}(nh),
\label{secondder}
\end{align}
and so on.

\section{A discretization of the homogeneous Euler equation} \label{sec4}

 In this section, as an application of the theory outlined above, we shall consider homogeneous linear equations of second order:
\begin{equation}\label{hom}
x^2 u''+axu' +b u=0,\quad a,b\in \mathbb{R}.
\end{equation}

We describe the discretization procedure step-by-step to illustrate it in a self-consistent way. A more abstract, formal approach based on category theory can also be implemented \cite{PTprep}.

\subsection{The construction of the model}
Given the differential equation \eqref{hom}, we substitute $x^k$ by $p_k(nh)$,  the derivatives by their discrete versions \eqref{firstder}, \eqref{secondder} and the usual product by the $*$ product. Explicitly, we have
\begin{equation}\label{lin}
  p_{2}(nh)* \sum_{k=0}^{n+2} \frac{k!\zeta_k}{(k-2)!}  p_{k-2}(nh)+a \,p_{1}(nh)* \sum_{k=0}^{n+1} \frac{k!\zeta_k}{(k-1)!}  p_{k-1}(nh)+  b\sum_{k=0}^{\infty}\zeta_kp_k(nh)=0,
\end{equation}
that is
\begin{equation} 
  \sum_{k=0}^{n}\left( \frac{1}{(k-2)!}   +   \frac{a}{(k-1)!}   +  \frac{b}{k!} \right)k!\zeta_kp_k(nh)=0,
\end{equation}
or, equivalently
\begin{equation} 
  \sum_{k=0}^{n}\left( k(k-1)+  a k   +  b\right) \zeta_kp_k(nh)= \sum_{k=0}^{n}\Lambda_k \zeta_kp_k(nh)=0,
\end{equation}
where
\begin{equation} \label{indicial}
\Lambda_k= k(k-1)+  a k   +  b 
\end{equation}
is the indicial polynomial of the Euler differential equation \eqref{hom}.

The difference equation for $u_k$ can be easily written using \eqref{pol} and \eqref{zeta}:
\begin{equation}
\sum_{k=0}^{n}\sum_{j=0}^{k}\frac{(-1)^{k-j}n!\Lambda_k }{j!(k-j)!(n-k)!} u_j=\sum_{k=0}^{n}\sum_{j=0}^{k}(-1)^{k-j}\Lambda_k\binom{n}{k}\binom{k}{j} u_j=0 \ .
\end{equation}
Changing the order of the sums, we obtain the relation
\begin{equation}
 \sum_{j=0}^{n}\left[\sum_{k=j}^{n}(-1)^{k-j}\Lambda_k \binom{n}{k}\binom{k}{j}\right]u_j =0
\end{equation}
which, using some combinatorial identities listed in the Appendix, becomes:
\begin{equation}\label{main}
\sum_{j=0}^{n}\binom{n}{j}\left[\sum_{k=j}^{n}(-1)^{k-j}\Lambda_{k} \binom{n-j}{k-j}\right]u_j = \sum_{j=0}^{n}\binom{n}{j}c_{nj}u_j =0.
\end{equation}
Here,
\begin{equation}
c_{nj}  =\sum_{k=j}^{n}(-1)^{k-j} \binom{n-j}{k-j}\Lambda_{k}= \sum_{k=0}^{n-j}(-1)^{k} \binom{n-j}{k}\Lambda_{k+j} .
\end{equation}
By using again the results in the Appendix, we get
\begin{equation}
\sum_{k=0}^{n-j}(-1)^{k}\binom{n-j}{k}\Lambda_{k+j}  = 
\sum_{k=0}^{n-j}(-1)^{k}\binom{n-j}{k}\left(k^2+(a+2j- 1 )k  +j^2+(a-1)j+b\right).
\end{equation}
This formula can be rewritten as
\begin{align}
& \sum_{k=0}^{n-j}(-1)^{k}\binom{n-j}{k}\Lambda_{k+j}= 
\sum_{k=0}^{n-j}(-1)^{k}\binom{n-j}{k}k^2+(a+2j-1)\sum_{k=0}^{n-j}(-1)^{k}\binom{n-j}{k}k  \notag \\  &\qquad  +\left(j^2+(a-1)j+b\right)\sum_{k=0}^{n-j}(-1)^{k}\binom{n-j}{k}\notag\\
  &\qquad =
2\delta_{n-j,2}-\delta_{n-j,1}-(a+2j-1)\delta_{n-j,1}    +\left(j^2+(a-1)j+b\right)\delta_{n-j,0} \ .
\end{align}
Finally,  for the coefficients of the difference equation \eqref{main}, we get
\begin{equation}
c_{nj}=2\delta_{n-2,j}-(a+2j )\delta_{n-1,j}+\left(j^2+(a-1)j+b\right)\delta_{n,j}\notag .
\end{equation}
Consequently, equation \eqref{main} reduces to a three-point difference equation. The only coefficients different from zero, for a given $n$,  are
\begin{equation}
c_{n,n-2}=2,\quad
c_{n,n-1}=-a-2(n-1),\quad
c_{nn}= n^2+(a-1)n+b \ .
\end{equation}
Therefore, our difference equation becomes:
\begin{equation} 
 \binom{n}{n-2}c_{n,n-2}u_{n-2}+ \binom{n}{n-1}c_{n,n-1}u_{n-1} + \binom{n}{n}c_{nn}u_n  = 0.
\end{equation}
Thus, we arrive at an interesting discrete model.
\begin{definition}
The difference equation
\begin{equation} \label{DE}
(n^2+(a-1)n+b)u_n - n(a+2n-2)u_{n-1} +n(n-1) u_{n-2} =0
\end{equation}
is said to be the \textit{discrete Euler equation}.
\end{definition}
\subsection{The continuous limit}
The difference equation \eqref{DE}, which has been obtained by an algorithmic, purely algebraic procedure, can be considered as the discrete version on the Rota algebra $(\mathcal{F}, +, *_{\Delta})$ of the continuous equation \eqref{hom}.

However, there is also a direct relation among eqs. \eqref{DE} and \eqref{hom}. More precisely, if $h\to 0$ and $nh$ remain finite with $nh\equiv x$, we can obtain the continuous limit of the difference equation by rewriting eq. \eqref{DE} as:
\begin{equation}
n(n-1)h^2 \frac{ u_n-2 u_{n-1}+ u_{n-2}}{h^2}+ a(nh)\frac{ u_n- u_{n-1}}{h}+bu_n   = 0.
\end{equation}
Then, in the limit when $n\to \infty$, our discrete model converts into the original differential equation
\begin{equation}
x^2 u''+axu' +b u=0.
\end{equation}
\subsection{Exact solutions of the discrete Euler equation}

As is well known, the space of solutions of the Euler equation is parametrized by the solutions $r_1$, $r_2$ of the indicial equation $\Lambda_r= r(r-1)+  a r   +  b =0$.
When $r_1, r_2 \in \mathbb{R}$, $r_1\neq r_2$, the Euler equation \eqref{hom} admits two independent solutions of the form $x^{r_i}$, $i=1,2$. When $r_1=r_2$, one of the two independent solutions is of this form, whereas the other contains a logarithmic singularity. When $r_1, r_2 \in \mathbb{C}$, we have two solutions expressed as a superposition of trigonometric functions of a logarithm. By construction, our method requires the existence of power series solutions for the original continuous model; thus, we will be able to determine two independent exact solutions of the form \eqref{FPE}  for the discrete Euler equation  when $r_1\neq r_2$ and one solution when the (real) roots coincide.
The case of complex roots would require an extension of the method, which is presently under investigation.

 Let us assume that $r\in \mathbb{N}$. This restriction on the coefficients $a$, $b$ in eq. \eqref{hom} ensures the existence of analytic solutions.

According to our methodology, to inherit exact solutions of the discrete equation \eqref{DE} from the continuous one, we expand the known solutions of eq. \eqref{hom} in terms of the basic sequence associated with the $\Delta$ derivative. By using eqs. \eqref{FPE}-\eqref{un}, we obtain the discrete function:
\begin{equation} \label{solDE}
u_n=p_r(n)=\frac{n!}{(n-r)!},\quad n\ge r.
\end{equation}
Let us show directly that this expression is an exact solution of the difference equation \eqref{DE}. By substituting it into the equation, we have
\begin{equation}
(n^2+(a-1)n+b)\frac{n!}{(n-r)!} - n(a+2n-2)\frac{(n-1)!}{(n-1-r)!}  +n(n-1) \frac{(n-2)!}{(n-2-r)!}  =0.
\end{equation}
After some  simplifications, we get the relation
\begin{equation}
(n^2+(a-1)n+b) - (a+2n-2)(n-r)  +  (n-r)(n-r-1)  =0,
\end{equation}
that reduces to
\begin{equation}
r(r-1)+ar+b=0
\end{equation}
which is an identity since $r$ is a root of the indicial polynomial.

\section{Future perspectives} \label{sec5}

This work is part of a research project concerning the discretization of ODEs and PDEs in the theoretical framework offered by the finite operator calculus and Rota algebras. In the case of the Euler equation, we have shown the effectiveness of our discretization procedure by deriving the new discrete model \eqref{DE}.  In particular, as we have shown, a family of exact solutions of this model  is inherited by construction from the continuous Euler equation.

For a large class of variable coefficients ODEs, several general results and different examples have been obtained in \cite{PTprep}. A discrete version of the classical Frobenius theorem for ODEs is also in order \cite{RRT}.

The discretization approach described in this article deserves further investigation. Among the related open problems, we wish to address the study of the symmetry groups admitted by the new discrete models obtained in this framework, in the spirit of Lie's approach to ODEs \cite{LR}, and the construction of appropriate boundary value problems. 

More generally, we plan to extend the discretization procedure based on Rota algebras to the case of both linear and nonlinear partial differential equations. Several preliminary results show the potential applicability of these ideas also to the construction of new integrable quantum models defined on a lattice \cite{FL}. This also paves the way for possible applications of the theory to discrete formulations of quantum gravity \cite{GP,RS}.

\appendix

\section*{Appendix: Some useful combinatorial identities}
The Newton binomial formula  holds for any $a\in\mathbb{C}$:

\begin{equation}\label{bin1}
\sum_{k=0}^{n}(-1)^{n-k}\binom{n}{k}a^k= (a-1)^{n}.
\end{equation}
When $a=1$, it reduces to
\begin{equation}\label{bin2}
\sum_{k=0}^{n}(-1)^{n-k}\binom{n}{k}= 0,\quad n>0.
\end{equation}
If $n=0$, for any $a$ we have
\begin{equation}\label{bin3}
\sum_{k=0}^{n}(-1)^{n-k}\binom{n}{k}a^k= 1 \ .
\end{equation}
Then, for $a=1$ and $n\in \mathbb{N}\cup \{0\}$ we have the identity
\begin{equation}\label{bin4}
\sum_{k=0}^{n}(-1)^{n-k}\binom{n}{k}=(-1)^n \delta_{n0}=\delta_{n0}.
\end{equation}
Taking formally the derivative with respect to $a$ in \eqref{bin1} and multiplying both sides by $a$, we obtain
\begin{equation}\label{bin5}
\sum_{k=0}^{n}(-1)^{n-k}\binom{n}{k}ka^{k}= na(a-1)^{n-1} \ .
\end{equation}
Thus, if $a=1$ we get
\begin{equation}\label{bin6}
\sum_{k=0}^{n}(-1)^{n-k}\binom{n}{k}k = 0,\quad n> 1.
\end{equation}
Again, putting $n=1$ in formula \eqref{bin6}, the r.h.s. equals 1, therefore
\begin{equation}\label{bin7}
\sum_{k=0}^{n}(-1)^{k}\binom{n}{k}k  =(-1)^n \delta_{n1}=-\delta_{n1}.
\end{equation}
Taking the second derivative in \eqref{bin1}, we deduce the relation
\begin{equation}\label{bin8}
\sum_{k=0}^{n}(-1)^{n-k}\binom{n}{k}k(k-1)a^{k-2}= n(n-1)(a-1)^{n-2},
\end{equation}
which can also be written as
\begin{equation}\label{bin9}
\sum_{k=0}^{n}(-1)^{n-k}\binom{n}{k}k^2a^{k}-\sum_{k=0}^{n}(-1)^{n-k}\binom{n}{k}ka^{k}= n(n-1)a^2(a-1)^{n-2}.
\end{equation}
Equivalently, we have
\begin{equation}\label{bin10}
\sum_{k=0}^{n}(-1)^{n-k}\binom{n}{k}k^2a^{k}=na(na-1)(a-1)^{n-2}.
\end{equation}
If $a=1$, the previous formula reduces to
\begin{equation}\label{bin11}
\sum_{k=0}^{n}(-1)^{n-k}\binom{n}{k}k^2=0,\quad n>2.
\end{equation}
However, if $n=1$, we have
\begin{equation}\label{bin12}
\sum_{k=0}^{n}(-1)^{k}\binom{n}{k}k^2 =(-1)^n \delta_{n1}=-\delta_{n1}
\end{equation}
and, if $n=2$
\begin{equation}\label{bin13}
\sum_{k=0}^{n}(-1)^{k}\binom{n}{k}k^2 =2(-1)^n \delta_{n2}=2\delta_{n2}.
\end{equation}
Thus, we obtain the formula
\begin{equation}\label{bin14}
\sum_{k=0}^{n}(-1)^{k}\binom{n}{k}k^2 =2\delta_{n2}-\delta_{n1}=\begin{cases}
0 ,& n=0\\
-1,& n=1\\
2,& n=2\\
0,& n>2
\end{cases} \ .
\end{equation}
In particular, for the case under study, we deduce
\begin{align}
\sum_{k=0}^{n}(-1)^{k}\binom{n}{k}\Lambda_k =&\sum_{k=0}^{n}(-1)^{k}\binom{n}{k}k^2 +(a-1)\sum_{k=0}^{n}(-1)^{k}\binom{n}{k}k +b\sum_{k=0}^{n}(-1)^{k}\binom{n}{k} \notag \\ =&2\delta_{n2}-\delta_{n1}-(a-1)\delta_{n1}+b\delta_{n0} 
= 2\delta_{n2}-a\delta_{n1}+b\delta_{n0}.
\end{align}

\subsection*{Acknowledgements}

The research of P. T. has been supported by the Severo Ochoa Programme for Centres of Excellence in R\&D
(CEX2019-000904-S), Ministerio de Ciencia, Innovaci\'{o}n y Universidades y Agencia Estatal de Investigaci\'on, Spain. M.A.R. and P.T. wish to thank the Grupo UCM - F\'isica Matem\'atica for financial support. P.T. is a member of the Gruppo Nazionale di Fisica Matematica (GNFM) of the Istituto Nazionale di Alta Matematica.

\label{lastpage}
\end{document}